\documentclass[preprint2]{aastex}
\slugcomment{Submitted to Astrophysical Journal Letters}
\lefthead{}
\righthead{Decoupling of Nested Bars}

\def \m3{{\rm Mark III}}

\def\gtorder{\mathrel{\raise.3ex\hbox{$>$}\mkern-14mu
    \lower0.6ex\hbox{$\sim$}}}
\def\ltorder{\mathrel{\raise.3ex\hbox{$<$}\mkern-14mu
    \lower0.6ex\hbox{$\sim$}}}

\begin{document}
\lefthead{Englmaier and Shlosman}
\righthead{Dynamical Decoupling of Nested Bars}

\title{DYNAMICAL DECOUPLING OF NESTED BARS:\\
SELF-GRAVITATING GASEOUS NUCLEAR BARS}

\author{Peter Englmaier}
\affil{Astronomisches Institut, Universit\"at Basel, Venisstrasse 7, 
CH-4102, Binningen, Switzerland\\
email: {\tt Peter.Englmaier@unibas.ch}}

\and 

\author{Isaac Shlosman} 
\affil{Department of Physics and Astronomy,
University of Kentucky, Lexington, KY 40506-0055, USA\\
email: {\tt shlosman@pa.uky.edu}}

\begin{abstract}
A substantial fraction of barred galaxies host additional nuclear bars which
tumble with pattern speeds exceeding those of the large-scale (primary) 
stellar bars. We have investigated the mechanism of formation and dynamical 
decoupling in such nested bars which include gaseous (secondary) nuclear bars 
within the full size galactic disks, hosting a double inner Lindblad resonance. 
Becoming increasingly massive and self-gravitating, the nuclear bars
lose internal (circulation) angular momentum to the primary bars and increase
their strength. Developing chaos within these bars triggers a rapid gas
collapse --- bar contraction. During this time period, the secondary bar
pattern speed $\Omega_{\rm s} \sim a^{-1}$, where $a$ stands for the bar size.
As a result, $\Omega_{\rm s}$ increases dramatically until a new equilibrium
is reached (if at all), while the gas specific angular momentum decreases ---
demonstrating the dynamical decoupling of nested bars. Viscosity, and
therefore the gas presence, appears to be a necessary condition for the
prograde decoupling of nested bars. This process maintains an inflow rate of 
$\sim 1~{\rm M_\odot~yr^{-1}}$ over $\sim 10^8$~yrs across the central 200~pc
and has important implications for fueling the nuclear starbursts and AGN.
\end{abstract}
\keywords{hydrodynamics --- galaxies: active --- galaxies: evolution --- 
galaxies: kinematics and dynamics --- galaxies: spiral ---  galaxies:
starburst}


\section{Introduction}

A large fraction, $\sim 1/3$, of barred galaxies host a secondary (nuclear) bar
in the central regions, in addition to the primary, large-scale stellar bar
(Laine et al. 2002; Erwin \& Sparke 2002).   Ironically, such nested bar
systems have been predicted theoretically, including their most intriguing
property --- the state of dynamical decoupling (Shlosman, Frank \& Begelman
1989; Shlosman, Begelman \& Frank 1990). Few examples of double barred
galaxies have been observed already in the 1970s, but have been classified as
having triaxial (elliptical) bulges rather than nuclear bars (de Vaucouleurs
1974; Sandage \& Brucato 1979; Kormendy 1982). Both stellar and mixed type bars 
(e.g., Buta \& Crocker 1993;  Shaw et  al. 1995;  Friedli et al. 1996; 
Jungwiert, Combes \& Axon 1997; Jogee, Kenney \& Smith 
1998; Knapen, Shlosman \& Peletier 2000), as well as gaseous nuclear bars 
(e.g., Ishizuki et al. 1990; Devereux, Kenney \& Young 1992; Forbes et al. 
1994; Mirabel et al. 1999; Kotilainen et al. 2000; Maiolino et al. 2000) have 
been detected. Random mutual orientation of the observed nested bars confirms 
that they spend a substantial fraction of their lifetime in the decoupled state
--- tumbling with different pattern speeds.

Numerical simulations have shown indeed that pattern speeds of the mixed
nuclear bars are substantially larger than those of the primary stellar bars
(e.g., Friedli \& Martinet 1993: Combes 1994; Heller \& Shlosman 1994).
However, purely stellar nuclear bars form only under special initial
conditions (e.g., Friedli \& Martinet 1993), while gaseous bars have been
claimed not to decouple (Wada \& Habe 1992; Combes 1994). A number 
of issues have never been resolved --- what triggers the dynamical 
decoupling of nested bars? Do purely gaseous self-gravitating bars form, and
can they decouple from the primary bars? In this Letter we use numerical 
simulations of self-gravitating gas in galactic disks to show that gaseous 
bars form and decouple as a result of stellar bar-driven inflow into the
central kpc. Furthermore, we study the physical reasons for this dynamical
runaway in nested bar systems (see also Shlosman 2003; and in
preparation).

\section{Numerical Modeling}

\begin{table}
\caption{Model Parameters}
\smallskip
\begin{tabular}{lccc} \hline\hline
Component & Mass ($10^{10}~{\rm M_\odot}$) & A (kpc)\\ \hline
halo        & 5     & 9   \\ 
bulge       & 0.4   & 0.2 \\
disk        & 2     & 3   \\
gas         & 0.05  & 3   \\ \hline 
\end{tabular}
\end{table}    

Cold gas forms a thin layer in disk galaxies. Because we are basically
interested in the dynamical evolution of this layer, we limit the simulations
to the disk midplane and use the updated 2-D version of our Smoothed Particle
Hydrodynamics (SPH) code (e.g., Heller \& Shlosman 1994). For test purposes we
also make use of the grid code ZEUS-2D (Stone \& Norman 1992). The initial
conditions have been tailored in order to focus on the role of central
resonances in disk galaxies by modifying the background analytical
potential of Miyamoto \& Nagai (1975), $ \Phi(r) = - {GM/\sqrt{r^2+A^2}}$,
using prescription from Englmaier \& Shlosman (2000). Here $M$ is the mass of 
a halo, bulge or a disk, and $A$ is the radial lengthscale, all given in
Table~1, and $r$ is the radius vector in the disk plane. Adjustments to
this potential allow us to specify the positions of axisymmetric resonances:
the inner inner and outer inner Lindblad resonances (IILR, OILR) and the
corotation. Positions of these resonances correspond to the solutions of
equation $\Omega-\kappa/2=\Omega_{\rm p}$, where $\Omega, \kappa$ and 
$\Omega_{\rm p}$ are angular velocity, epicyclic frequency and primary bar
pattern speed. For the models presented here, those are fixed at 
$r=0.5$~kpc, 2~kpc and 6~kpc respectively. The primary stellar bar is given by 
the Ferrers' (1877) potential with $n=1$ and rotates with a prescribed pattern 
speed $\Omega_{\rm p}=25.46~{\rm Gyr^{-1}}$. Its mass, the semi-major ($a$)
and semi-minor ($b$) axes are $4\times 10^9~{\rm M_\odot}$, 5~kpc and 1.25~kpc,
respectively. The axisymmetric contribution from the primary bar mass is
included in the initial conditions. The non-axisymmetric part is brought up
gradually in order to avoid transients. The number of SPH particles has been
varied between $N=10^4$ and $4\times 10^5$ and the minimal gravitational
softening has been fixed at 150~pc. A large number of models have been run
with increasing gas contents. Here we only discuss a representative model
showing a nested bar descoupling, with the gas mass $5\times 10^8~{\rm
M_\odot}$ 
within the corotation, which
corresponds to $\sim 2.7\%$ of the total mass within 3~kpc --- the initial
radial scalelength of the isothermal gas. The gas sound speed is $10~{\rm
km~s^{-1}}$. 

\section{Results}

\begin{figure*}[ht!!!!!!]
\vbox to3.6in{\rule{0pt}{3.6in}}
\includegraphics{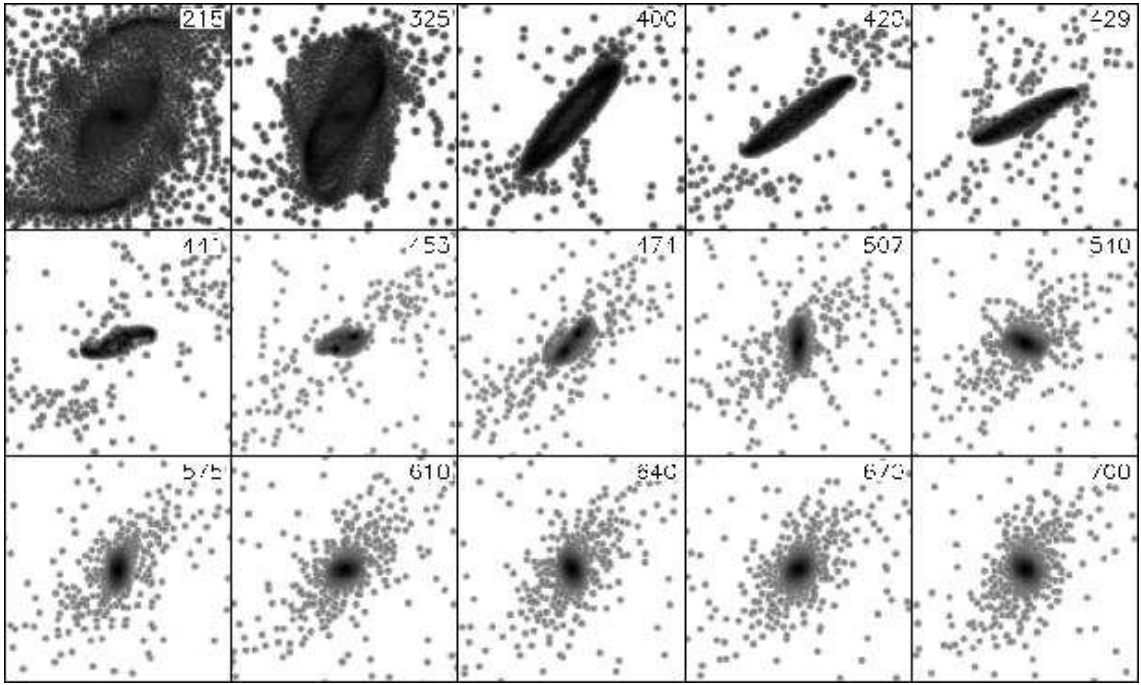}
\caption{Gas evolution in the central 3~kpc$\times$3~kpc part of a barred
galaxy in the frame of the primary bar (Animation Seq.~1). Rotation is
anti-clockwise and the primary bar is horizontal. The time in Myrs is shown in
the right upper corners (see text for more details). 
}
\end{figure*}

As the large-scale bar is brought up gradually, the gas
within the corotation radius forms a pair of trailing offset shocks and falls
inwards losing its angular momentum via gravitational torques induced by the
primary bar (see Fig.~1 and Animation Sequence~1). At around $\sim 1$~kpc, the
shocks curve azimuthally, and the gas flow is shock-focused into more circular
orbits attempting to form an ovally-shaped nuclear ring. The gas inside the
central kpc is losing its angular momentum forming a secondary bar which leads
the primary bar by a constant angle $\sim 50^\circ-60^\circ$ and a pattern
speed $\Omega_{\rm s}= \Omega_{\rm p}$, and has a flat
density distribution. A pair of leading shocks form between this nuclear bar
and the ring. Almost immediately the ring fragments and gradually merges with
the nuclear bar. At around $t\sim 0.35-0.4$~Gyr
this secondary bar with a semimajor axis of 1~kpc and ellipticity $\epsilon
\sim 0.65$ decouples in the {\it retrograde} direction, in the primary bar
frame (still prograde in the inertial frame). One can follow up the subsequent
evolution in one of the three frames of reference: inertial, primary or
secondary. We mostly employ the first two.

This unusual behavior of the gaseous bar is a result of gravitational torques
from the primary bar (Heller, Shlosman \& Englmaier 2001; see also below).
Its swing towards the position angle of the primary bar is
accompanied by a loss of internal (i.e., circulation) and precession (tumbling)
angular momenta in the gas. The bar ellipticity increases to $\sim 0.85$
(Fig.~2) and $\Omega_{\rm s}$ drops to $\sim 12~{\rm Gyr^{-1}}$ (Fig.~3) in the
inertial frame. Next, the gaseous bar reverses its tumbling to the
{\it prograde} one (in the primary frame) and quickly shrinks to much smaller
radii.
This reversal coincides with a dramatic gas inflow (Fig.~4). The gas density
distribution in the secondary bar becomes centrally peaked. The
avalanche inflow clearly leads to a period of an unstable bar shape, $\sim
0.1$~Gyr long. The bar apparently fragments during the collapse
and two fragments preserve their identity and store some of the 
circulation angular momentum during this time (e.g., $t=0.474$~Gyr in Fig.~1).
The
final value of $\Omega_{\rm s}\sim 85~{\rm Gyr^{-1}}$ is achieved at $t\sim
0.6$~Gyr with some superposed bar length oscillations which are well
correlated with the mutual orientation of both bars. 

\begin{figure}[ht!!!!!!]
\vbox to2.9in{\rule{0pt}{3.0in}}
\includegraphics{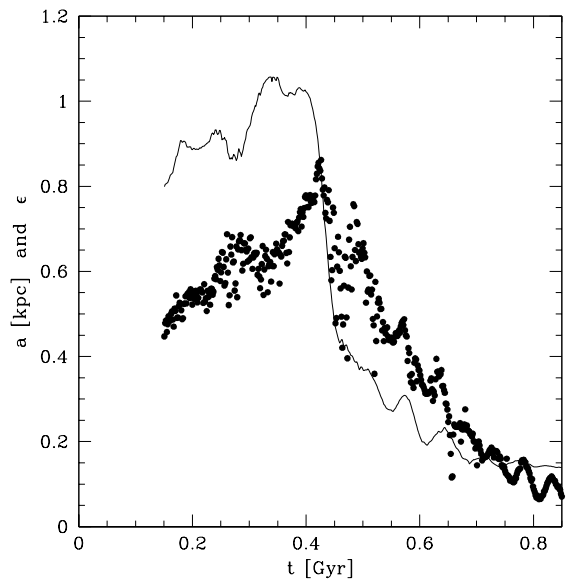}
\caption{Evolution of ellipticity, $\epsilon=1-b/a$ (dotted line), and
semimajor axis, $a$ (solid line) of the secondary bar. 
}
\end{figure}
\begin{figure}[ht!!!!!!]
\vbox to2.9in{\rule{0pt}{3.0in}}
\includegraphics{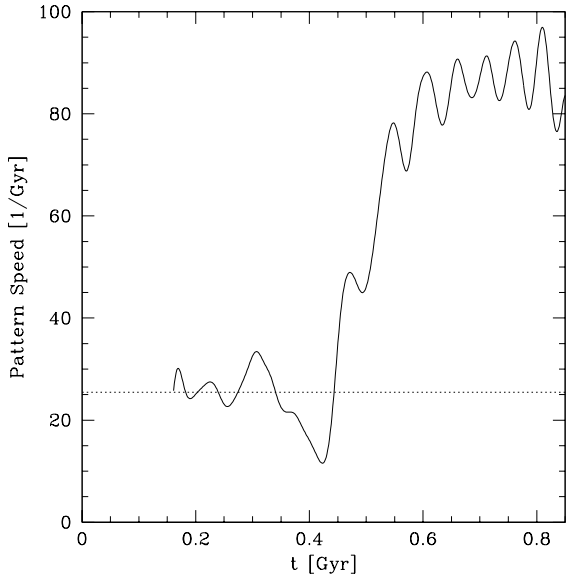}
\caption{Pattern speeds of primary ($\Omega_{\rm p}$, dotted line) and
secondary ($\Omega_{\rm s}$, solid line) bars in the inertial frame. 
}
\end{figure}
\begin{figure}[ht!!!!!!]
\vbox to2.9in{\rule{0pt}{3.0in}}
\includegraphics{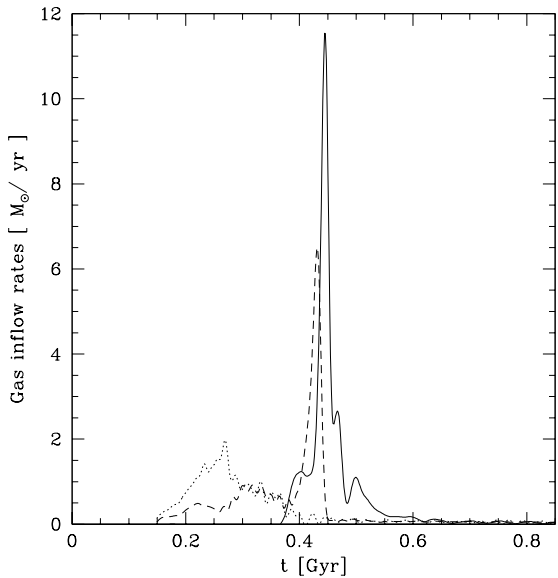}
\caption{Evolution of gas inflow rates across the central 175~pc (solid line),
600~pc (dashed line) and 1~kpc (dotted line). The peak at 0.45~Gyr is
unresolved.
}
\end{figure}

The observed pattern speed reversal (i.e., speedup in the inertial frame)
represents the dynamical runaway (decoupling) of the gaseous bar from the
primary bar. The secondary pattern speed has first decreased from $\Omega_{\rm
s} = \Omega_{\rm p}$ to its minimal value of $\sim 12~{\rm Gyr^{-1}}$ and then 
increased to the maximal $\sim 85~{\rm Gyr^{-1}}\sim 3.4\Omega_{\rm p}$ where 
it has stabilized (Fig.~3). By $t\sim 0.7$~Gyr the bar has shrank to $\sim
150$~pc.
This final bar size is the result of a limiting gravitational softening
(Section~2). Fig.~4 shows the gas mass inflow rates across a number of
characteristic radii. The time delays between the peak rates confirm that
the avalanche inflow propagates from larger to smaller spatial scales.
The mass ratio of secondary-to-primary bar, $f$, toward
the end of this evolution is $\sim 0.1$, and the gaseous bar contains about
$16\%$ of the total mass within its radius.
 
Next, we analyze this gas evolution and address the focal questions ---
what exactly triggers the secondary bar decoupling process and what determines
the final secondary pattern speed. We only focus on the
essential details here. Some of the
observed features in this model do not seem to be of a principal importance.
For example, the retrograde swing of the gaseous bar, although is interesting
and maybe relevant in some systems, is not crucial for the subsequent
prograde decoupling as other models show (in preparation). On the other hand,
we find that changes in the gas density and bar size lead directly to the
increase in the central mass concentration in the bar, thus amplifying the
self-gravitational effects in the gas, affecting dynamics of the system, and
triggering the observed runaway. 

First, when the secondary bar forms, it extends to the radius of the 
Max($\Omega-\kappa/2$) curve, at $\sim 1$~kpc. This is also the radius where 
the nuclear ring attempts to form. The radial density
profile in the gaseous bar is flat and even centrally depressed, because the 
IILR not only prevents the gas 
inflow to within 500~pc of the center, but also applies torques which expell 
some of the gas across the IILR. The gas response between the ILRs is 
$\sim 50-60^\circ$ out of phase with the primary bar\footnote{The so-called
periodic $x_2$ orbits in the notation of Contopoulos \& Papayannopoulos (1980)
are oriented at $90^\circ$ to the bar and serve as attractors for the gas
motion. The viscous torques prevent the gas response from being completely
aligned with the $x_2$ orbits} defining the initial position angle of the
secondary bar. The Jacobi energies, $E_{\rm J}$ (e.g., Binney \& Tremaine
1987), of the gas are distributed around the IILR as a Gaussian with $\Delta 
E_{\rm J}/E_{\rm J}\sim 5\%-10\%$ (e.g., Fig.~4 of Heller et al. 2001). 

Second, two important developments accompany this evolution: the gaseous bar
mass increases due to the continuous inflow, and the specific {\it internal}
angular
momentum (i.e., circulation) in the bar decreases due to the ongoing torquing
from the primary bar (note that the secondary bar is positioned in the first
and third quadrants). Loss of internal circulation increases the bar
ellipticity by diminishing its minor axis, thus forcing the gas inwards across
the IILR (i.e., as measured by $E_{\rm J}$ distribution of gas particles).
When $\sim 50\%$ of the gas mass crosses the IILR, it does not find any
supporting periodic orbits --- these are aligned with the primary bar inside
the IILR. The primary bar then torques the gas, which, because it leads the
bar by less than $90^\circ$, swings in the retrograde fashion towards the main
bar. Fig.~3 clearly displays this $\Omega_{\rm s}$ slowdown in the inertial
frame.

Third, increasing $\epsilon$ (i.e., bar strength) to $\gtorder 0.8$ in the
stellar bars generates a substantial chaos (e.g., Udry \& Pfenniger 1988;
Martinez-Valpuesta \& Shlosman 2004) ---  the self-gravity is necessary for
this. For a gaseous bar, this leads to a shock dissipation and to a dramatic
gas inflow. El-Zant  \& Shlosman (2003) have studied the stability of
nested bar systems using the Liapunov exponents and find that this system
has a narrow ``window'' for a stable co-existence --- otherwise a mutually
excited chaos develops. If $f\ltorder 0.01$, the secondary bars cannot exist in
the decoupled phase as no orbits aligned with these bars have been found. On
the other hand, when $f\gtorder 0.15$, the primary bar becomes increasingly
chaotic at the bar-bar interface. The value $f\sim
0.1$ obtained here fits comfortably within the allowed range. 

The main effect of intrinsic dissipation is the sudden increase in the central
mass concentration and in the gas self-gravity. This is
nicely exhibited by the radius of Max($\Omega-\kappa/2$) which is dragged
inwards by the collapsing gas from $\sim 1$~kpc. In fact, the secondary bar
length follows exactly this radius, which represents the position
of the ILRs to a large degree. This is not surprising, as these resonances are
main generators of chaos in disk galaxies. Because the gas cannot reside on
intersecting orbits, gaseous bars are expected to be limited in size to within
their ILRs, if the latter exist.  

Although the gas in the model contributes only $\sim 2\%$ of the total
mass within the primary bar corotation at 6~kpc, it alters substantially the
position of the IILR at later times, as the gas fraction of the total mass
within the
secondary bar radius ends up at $\sim 16\%$. This is verified through the
appearance of a characteristic orbital family in the nonlinear orbit diagram
(in preparation), but even the linear (epicyclic) diagrams exhibit this
clearly.
 
Overall, the gaseous bar shrinks by a factor of $\sim 7$ from the decoupling
moment and its pattern speed $\Omega_{\rm s}$ increases by the same amount 
(Figs.~2, 3). Their product, $\Omega_{\rm s}a \sim const.$, varies by $\pm
22\%$
only. 
This corresponds to a decrease in the tumbling angular momentum of the bar, 
$\sim \Omega_{\rm s}a^2$, by a factor of $\sim 7$. Qualitatively
this means that the gaseous bar is speeding up when contracting, and is losing
some of its angular momentum to the primary bar torques. To verify that this 
effect is driven by the gas self-gravity, we have evolved the same model
neglecting the gas gravity. The evolution diverges at the retrograde swing ---
the non-self-gravitating bars do not generate additional chaos and
dissipation, and do not decouple in the prograde direction. Requirement
for dissipation underscores that a sufficient viscosity, and, therefore, a 
presence of the gas is a necessary condition for prograde decoupling of nested 
bars.

We note, that approximate constancy of $\Omega_{\rm s}a$ for the decoupling bar 
discussed above has some analogy with the secular evolution of stellar bars.
Those are known to extend to their corotation (e.g., Athanassoula 1992). 
Because the corotation radius typically lies in the flat part of the galactic 
rotation curve and so $\Omega\sim r^{-1}$, this leads $\Omega_pa\sim const.$
One can show that the same relation can be worked out in the epicyclic
approximation for decoupling gaseous bars as well, on the assumption that
their size is limited by the radius of the IILR --- which is supported by
our simulations. 

An interesting question is to what degree both bars in the decoupled phase
exchange energies via mechanism known as a mode coupling (e.g., Sygnet et 
al. 1988; Massett \&
Tagger 1997). The corotation radius corresponding to the final $\Omega_{\rm s}$ 
is positioned close enough to the OILR of the primary bar, and their
nonlinear interaction here is not out of question.
However, mode coupling does not seem to contribute to the nested bar
decoupling itself. It can of course play an important role in locking these two
modes in the resonance interaction when both bars are fully developed. This 
issue will be addressed elsewhere. 
 
We note two more comments. First, the initial retrograde (in the frame of the
primary bar) swing of the secondary bar does not seem to be of a
principal importance. When the gas mass is increased just by 20\%, the bar
decouples directly into the prograde direction. Second, the model gaseous bar
is expectedly triggered by the primary bar. In the control run with the same
mass distribution but without the non-axisymmetric contribution from the
large-scale bar, the nuclear bar did not develop and the central kpc remained
nearly axisymmetric.  

In summary, in a fully developed galactic disk with a double ILR, a secondary
gaseous bar forms in response to the gas inflow along the stellar bar.
Initially corotating with the primary bar, this gaseous bar extends to radii
where nuclear rings are known to form. Subsequent mass inflow and loss of
circulation angular momentum to the stellar bar via gravitational torques
strengthens this bar --- a process which leads to development of chaotic
orbits within the bar and, unlike in purely stellar bars, to a rapid gas
infall of $\sim 1~{\rm M_\odot~yr^{-1}}$ over $\sim 10^8$~yrs across the
central 200~pc, as gas cannot populate intersecting orbits. As the bar pattern 
speed appears to be inversely proportional
to bar's shrinking size, the pattern speed increases manyfold, triggering
a runaway dynamical decoupling of the gaseous bar from the stellar bar. The
gaseous bars extend only to their ILRs, unlike their stellar counterparts.
Although numerical
simulations of this avalanche inflow necesarily are limited by a finite
dynamic range which cuts off the inflow, it can in principle proceed to much
smaller scales and has implications for fueling nuclear starbursts and AGN. The 
outcome of this process depends on a number of physical processes which are not 
included in numerical simulations here, such as a runaway star formation and 
the degree of clumpiness of the interstellar matter. 


\acknowledgments
We thank Amr El-Zant, Ingo Berentzen, Clayton Heller and Barbara Pichardo for
fruitful discussions, technical assistance and comments on manuscript. I.S. is
partially supported by
NASA/ATP/LTSA grants NAG5-10823, 5-13063, by NSF AST-0206251, and by HST
AR-09546 and 10284 from STScI, which is operated by AURA for NASA, under
NAS5-26555.

{}




\end{document}